\documentclass[fleqn,usenatbib]{mnras}

\usepackage{newtxtext,newtxmath}

\usepackage[T1]{fontenc}

\DeclareRobustCommand{\VAN}[3]{#2}
\let\VANthebibliography\thebibliography
\def\thebibliography{\DeclareRobustCommand{\VAN}[3]{##3}\VANthebibliography}

\usepackage{graphicx}	
\usepackage{amsmath}	

\title[RSGs on the brink of core-collapse]{Explosion Imminent: the appearance of Red Supergiants at the point of core-collapse}

\author[Davies, Plez \& Petrault]{
Ben Davies$^{1}$\thanks{b.davies@ljmu.ac.uk}, Bertrand Plez$^{2}$ and Mike Petrault$^{2}$
\\
$^{1}$Astrophysics Research Institute, Liverpool John Moores 
University, Liverpool Science Park ic2, mv 146 Brownlow Hill, Liverpool, L3 5RF, UK.\\
$^{2}$Laboratoire Univers et Particules de Montpellier, Univ Montpellier, CNRS, Montpellier, France.\\
}

\date{Accepted XXX. Received YYY; in original form ZZZ}

\pubyear{2021}

\begin{document}
\label{firstpage}
\pagerange{\pageref{firstpage}--\pageref{lastpage}}
\maketitle

\begin{abstract}
From the early radiation of type II-P supernovae (SNe), it has been claimed that the majority of their red supergiant (RSG) progenitors are enshrouded by large amounts of circumstellar material (CSM) at the point of explosion. The inferred density of this CSM is orders of magnitude above that seen around RSGs in the field, and is therefore indicative of a short phase of elevated mass-loss prior to explosion. It is not known over what timescale this material gets there: is it formed over several decades by a `superwind' with mass-loss rate $\dot{M} \sim10^{-3}\,{\rm M_\odot\,yr^{-1}}$; or is it formed in less than a year by a brief `outburst' with $\dot{M}\sim10^{-1}\,{\rm M_\odot\,yr^{-1}}$? In this paper, we simulate spectra for RSGs undergoing such mass-loss events, and demonstrate that in either scenario the CSM suppresses the optical flux by over a factor of 100, and that of the near-IR by a factor of 10. We argue that the `superwind' model can be excluded as it causes the progenitor to be heavily obscured for decades before explosion, and is strongly at odds with observations of II-P progenitors taken within 10 years of core-collapse. Instead, our results favour abrupt outbursts $<$1 year before explosion as the explanation for the early optical radiation of II-P SNe. We therefore predict that RSGs will undergo dramatic photometric variability in the optical and infrared in the weeks-to-months before core-collapse. 
\end{abstract}

\begin{keywords}
keyword1 -- keyword2 -- keyword3
\end{keywords}

\def\ga{\mathrel{\hbox{\rlap{\hbox{\lower4pt\hbox{$\sim$}}}\hbox{$>$}}}}
\def\la{\mathrel{\hbox{\rlap{\hbox{\lower4pt\hbox{$\sim$}}}\hbox{$<$}}}}
\def\msunyr{M\mbox{$_{\normalsize\odot}$}\rm{yr}$^{-1}$}
\def\msun{$M$\mbox{$_{\normalsize\odot}$}}
\def\zsun{$Z$\mbox{$_{\normalsize\odot}$}}
\def\rsun{$R$\mbox{$_{\normalsize\odot}$}}
\def\rstar{$R_\star$}
\def\minit{$M_{\rm init}$}
\def\lsun{L\mbox{$_{\normalsize\odot}$}}
\def\mdot{$\dot{M}$}
\def\logmdot{$\log(\dot{M}/{\rm M_\odot\,yr^{-1}})$}
\def\mdotdj{$\dot{M}_{\rm dJ}$}
\def\lbol{$L$\mbox{$_{\rm bol}$}}
\def\kms{\,km~s$^{-1}$}
\def\EW{$W_{\lambda}$}
\def\arcsec{$^{\prime \prime}$}
\def\arcmin{$^{\prime}$}
\def\teff{$T_{\rm eff}$}
\def\Teff{$T_{\rm eff}$}
\def\logg{$\log g$}
\def\logz{$\log Z$}
\def\logl{$\log (L/L_\odot)$}
\def\vdisp{$v_{\rm disp}$}
\def\vinf{$v_{\infty}$}
\def\bcv{{\it BC$_V$}}
\def\bci{{\it BC$_I$}}
\def\bck{{\it BC$_K$}}
\def\lmax{$L_{\rm max}$}
\def\um{$\mu$m}
\def\chisq{$\chi^{2}$}
\def\AV{$A_{V}$}
\def\hminus{H$^{-}$}
\def\Hminus{H$^{-}$}
\def\ebmv{$E(B-V)$}
\def\mdyn{$M_{\rm dyn}$}
\def\mphot{$M_{\rm phot}$}
\def\cnterm{[C/N]$_{\rm term}$}
\newcommand{\fig}[1]{Fig.\ \ref{#1}}
\newcommand{\Fig}[1]{Figure \ref{#1}}
\newcommand{\newtext}[1]{{\bf #1}}
\newcommand{\nntext}[1]{{#1}}

\section{Introduction}
For the few dozen II-P supernovae (SNe) that have occured within 30Mpc, it has been possible to identify the progenitor star in high-resolution imaging taken 1-10 years before it exploded. In the cases where a progenitor is identified, the star is always inferred to be both luminous ($> 10^{4.5}$\lsun) and with colours consistent with spectral type M \citep{smartt04,Smartt09,Smartt15,Davies-Beasor18,Davies-Beasor20}. This evidence therefore strongly points towards Red Supergiants (RSGs) as being the direct progenitors for the most common type of core-collapse SN.

Whilst pre-explosion imaging can tell us what the star looked like within a decade of the SN, the SN radiation itself can give us hints about what the star looked like at the moment the core collapsed. In particular, in the first few days after core-collapse as the explosion `breaks out' of the star's optically thick surface layers we may learn about the physical conditions at the star's upper atmosphere and inner circumstellar material (CSM). Spectroscopy taken within a day or so of shock breakout reveal narrow emission lines which originate in the dense CSM, but which disappear quickly as the CSM is overrun by the blast-wave \citep[e.g.][]{Yaron17,Jacobson-Galan22}. In both works cited, analysis of this `flash' spectroscopy has suggested circumstellar densities of between 10$^{-13}$ to $10^{-14}$g\,cm$^{-3}$ at a few $\times 10^{14}$cm, which is orders of magnitude more dense than, for example, the CSM around Betelgeuse \citep{Harper01}. Subsequent similar studies of larger samples of SNe find that the fraction of objects demonstrating this behaviour increases the closer the SN is observed to explosion \citep{Khazov16,Bruch21}. Indeed, the latter study found 6 out of 10 objects observed within 2 days displayed flash-ionised spectral features. This hints that perhaps the majority of II-P SNe might exhibit this evidence for dense CSM around the progenitor as long as they are observed early enough. 

Corroborating evidence for dense CSM around II-P progenitors comes from their early-time light curves. If the CSM is optically thick, it delays the breakout of the shock, allowing the explosion to cool further before it emerges. This leads to a shorter timescale for the SN radiation to shift into the optical bands, causing sharper rise-times in the optical lightcurve \citep{Moriya17,Morozova17}. Hydrodynamical modelling of the early optical lightcurve, and the steepness of the initial rise, can then be used to estimate the radius/density of the shock breakout location. Early work on single SNe indicated  densities $>10^{-10}$g\,cm$^{-3}$ within $\sim$0.5\rstar\ of the stellar surface \citep{Moriya17,Morozova17,Dessart17}. These densities are many orders of magnitude higher than those inferred around typical RSGs, or even RSGs with very high mass-loss rates \citep[see Fig.\ 1 of ][]{Davies-Plez21}. Later work on larger samples of objects suggests that evidence for dense CSM is seen around 70-90\% of II-P SNe studied \citep{Foerster18,Morozova18}. 

The open question now is how this pre-SN CSM was formed, and a clue to answering this question will come from establishing over what timescale the CSM forms. The models put forward to explain this existence of the CSM fall into roughly two camps, which we will refer to as the `outburst' and `superwind' scenarios. Each of these scenarios will be discussed in more detail in Sect.\ \ref{sec:models}, but briefly they can be differentiated by the timescale over which they form the CSM. In the `outburst' scenario, the CSM is the product of a very high mass-loss rate event ($\dot{M} \sim 0.1$\msunyr) lasting of order a year \citep[e.g.][]{Morozova18}. In the `superwind' scenario, the CSM is the product of a steady-state outflow with a very high mass-loss rate, which operates for $\ga$100 years prior to explosion \citep[e.g.][]{Foerster18}. A third possibility is discussed in \citet{Dessart17}, in which the atmospheric scale-height is puffed up by factors of 10 to 30, with a wind stitched on top. These models generally have lower CSM masses than the `superwind' or 'outburst' models, but have greater CSM densities within $\sim$2\rstar ($\times 10^{3-5}$ higher than typical RSGs). No timescale for the generation of the CSM is provided for this latter scenario. 

Hydrodynamical modelling of the supernova's early-time lightcurve alone is not capable of discriminating between the two scenarios. However, since each scenario makes a different predictions about how the progenitor star should look in the years before SN, they can be contrasted by simulating how the progenitor should look in the decade or so before explosion. In this paper, we will construct models replicating each scenario\footnote{\citet{Dessart17} do not provide timescales for how long it takes to generate the CSM in their models, so at this time we do not explore these models' synthetic spectra. In this current paper, we investigate only the \citet{Foerster18} and \citet{Morozova20} models.} (Sect. \ref{sec:models}), and use them to compute synthetic spectra for each of the two scenarios described above (Sect\ \ref{sec:spectra}). We will then compare our results to observations of progenitors in the literature (Sect.\ \ref{sec:disc}). We conclude in Sect.\ \ref{sec:conc}


\section{Description of pre-SN models} \label{sec:models}

\begin{figure}
    \centering
    \includegraphics[width=8.5cm]{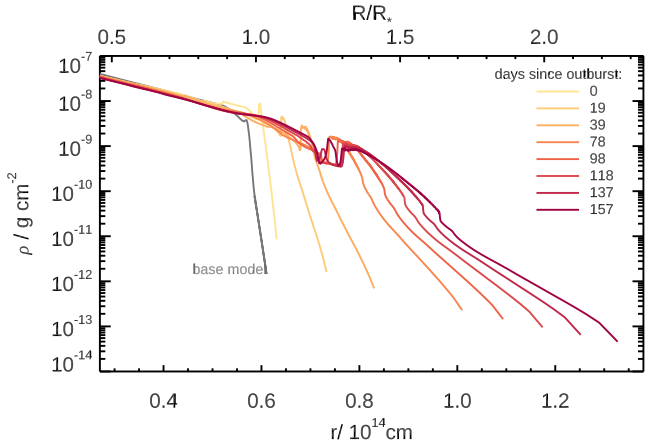}
    \caption{Density profiles of the outburst models of M20 as a function of time since the shockwave pierced the progenitor surface, which we define here as day 0 (defined as day 140 in M20). The last model, at 157 days since the emergence of the outburst, is that determined by to be the best-fitting pre-explosion density profile. }
    \label{fig:dens_mz}
\end{figure}

\subsection{`Outburst' models}
For the `outburst' model, we refer to the work of \citet[][ the latter two papers hereafter referred to as M18 and M20]{Morozova17,Morozova18,Morozova20}. In M18, it was argued that the early-time lightcurves of a sample of SNe suggested that in the majority of cases an excess of 0.1\msun\ of CSM was confined within $\sim 10^{14}$cm, which is approximately 2$R_\star$ for a typical RSG. To achieve this mass of CSM, these authors suggested that a wind of mass-loss rate \mdot$\sim$0.1\msunyr\ (for a putative wind speed of 10\kms{}) would be required. Such mass-loss rates are far many orders of magnitude above those of typical massive star winds, and are associated more with eruptive outbursts (e.g. $\eta$~Car) rather than quiescent winds. The sheer scale of such outbursts suggests that they would require a large, subsurface energy deposition rather than be opacity-driven as in a normal massive star wind. 

In a follow-up paper, M20 suggested a potential source of such an energy release could be wave-heating associated with late-stage nuclear burning. To simulate such an event, these authors took the pre-SN 15\msun\ model of \citet{Sukhbold18}, manually deposited between $10^{46}-10^{47}$ ergs of energy at the base of the convective envelope, and computed the evolution of the stellar structure as the shockwave propagated outwards through the envelope. Once the shock had reached the surface (which we define in this study as the beginning of the `outburst') and expanded the outer layers, M20 took a number of models at different timesteps, manually exploded them, and computed the resulting lightcurve. The conclusion of the paper was that the lightcurves of II-P SNe were best matched by outbursts which pierce the stellar surface $\sim$160 days before explosion, at which time around 1\msun\ of CSM has built up at radii above the original photosphere.

The density profiles as a function of time after the shock has penetrated the stellar surface are shown in \fig{fig:dens_mz}\footnote{We define day 0 as the first model in which the outburst has pierced the stellar surface. This is 140 days after the energy was deposited at the base of the convective envelope, and so this model is defined as day 140 in M20}. If we define the circumstellar mass $M_{\rm CSM}$ as that mass located above the original progenitor surface, the mass at +160 days is extremely high -- over 1\msun, an order of magnitude above those inferred in M18. To generate this much CSM in such a short time, this implies mass flow rates $>$1\msunyr. 


\subsection{`Superwind' models}
In \citet{Moriya17} it was first suggested that a wind with a mass-loss rate orders of magnitude below that inferred by \citet{Morozova16} may also be able to explain the early-time SN radiation, provided the wind had an extended acceleration zone. At fixed \mdot, the circumstellar density is inversely proportional to the outflow speed, and so a wind that expands very slowly until it reaches large radii will be much more dense in the inner regions. Thus, a slowly-accelerating wind with a much lower \mdot\ may have the same CSM density in the inner wind as a higher \mdot\ outburst which is accelerated more steeply.

However, even with such a shallow acceleration zone, mass-loss rates of order $10^{-3}$\msunyr\ are still required: orders of magnitude larger than those of typical RSGs, and a factor of $\sim$100 higher than even the most extreme objects \citep[e.g.][]{Beasor-Davies18,Beasor20,Humphreys20}. Furthermore, with a much reduced \mdot\ with respect to the `outburst' scenario, such a wind would need to operate for 100s of years prior to SN in order to build up the CSM necessary to reproduce the early-time lightcurves (cf.\ 100s of days for the `outburst' model).

\begin{figure}
    \centering
    \includegraphics[width=\columnwidth]{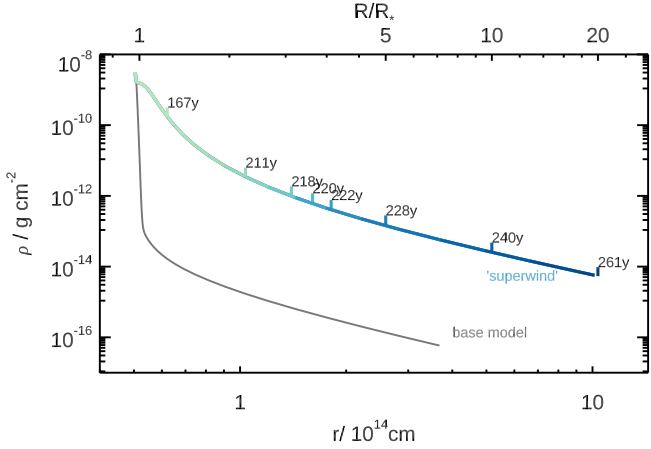}
    \caption{Density profile of the `superwind' model. The extent of the superwind's outer radius as a function of years since its initiation is indicated by the data labels. }
    \label{fig:dens_f18}
\end{figure}

\begin{figure*}
    \centering
    \includegraphics[width=17cm]{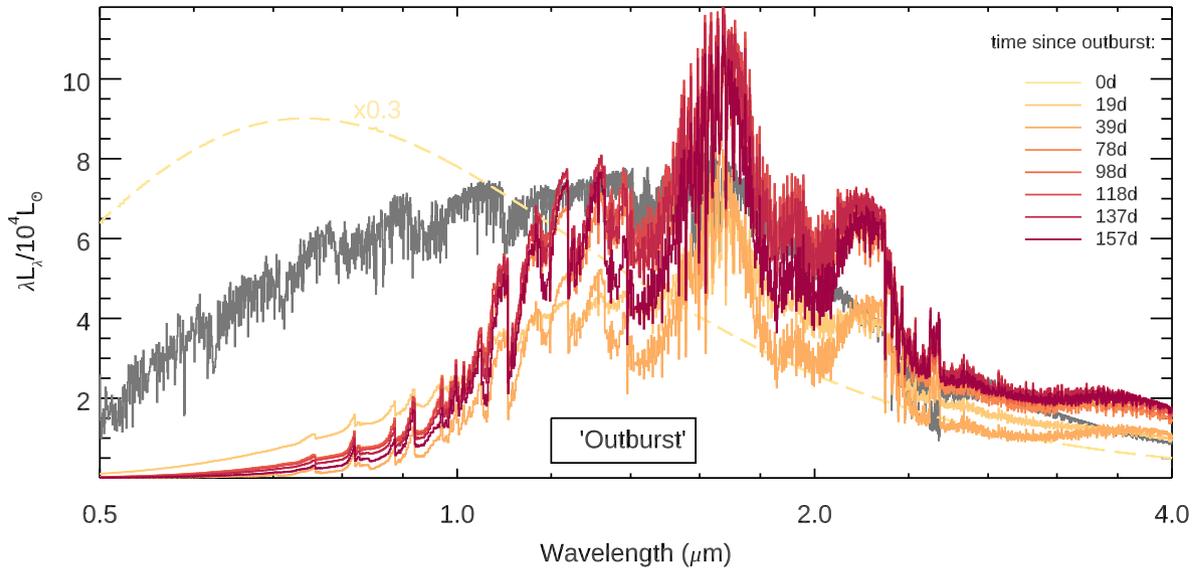}
    \caption{Simulated spectra of the outburst models, as a function of time since the initial outburst. All models have been smoothed to a constant resolving power of $\lambda/\Delta \lambda = 1000$. The earliest model has been scaled down by a factor of 0.3 for clarity. {\bf The spectrum of the base model is shown in grey.}}
    \label{fig:spectra-mz}
\end{figure*}
\begin{figure*}
    \centering
    \includegraphics[width=17cm]{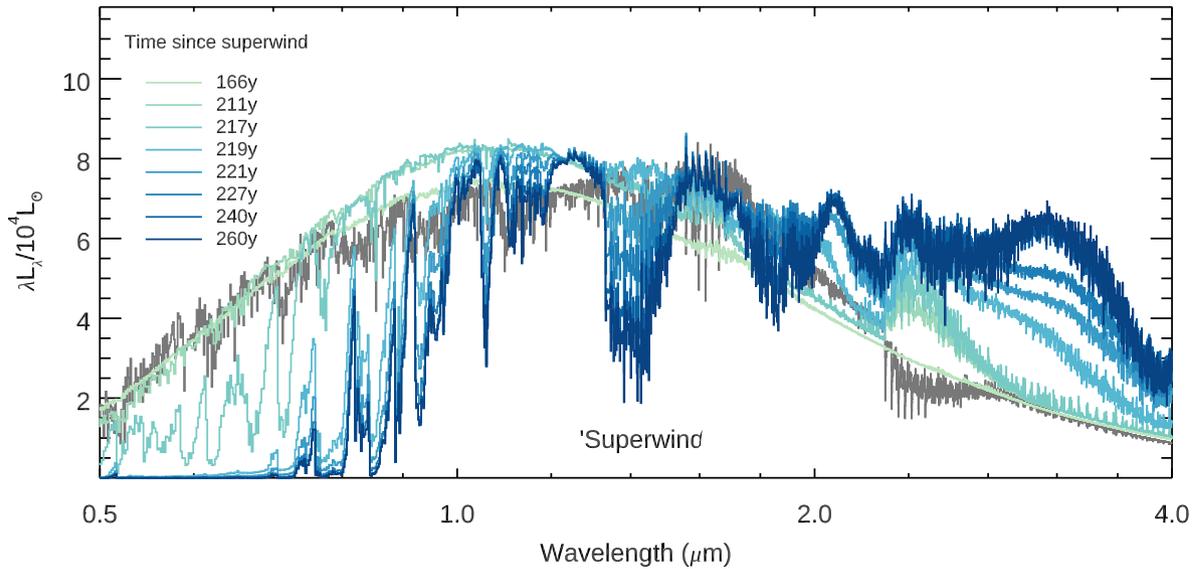}
    \caption{Simulated spectra of the `superwind' models, once the wind has extended to the radii indicated by the legend. {\bf The spectrum of the base model is shown in grey.} }
    \label{fig:spectra-f18}
\end{figure*}

\subsubsection{Construction of the `superwind' models}
Following \citet[][]{Davies-Plez21}, for our base model we begin with a MARCS model atmosphere of effective temperature \teff=3800K, gravity \logg=0.0, Solar-scaled abundances, and microturbulence $\xi=4$\kms\footnote{As we will see later, the details of the stellar atmosphere are relatively unimportant as it is almost completely masked by the CSM.}. The luminosity $L = 10^5$\lsun, and the radius $R_\star$=730\rsun. To this atmosphere we add a nominal ambient wind with \mdot=$10^{-7}$\msunyr, following the same procedure as \citet{Davies-Plez21}.

To this {\bf base }model we add CSM with a density profile $\rho$ varying with radius $r$ according to the mass continuity equation, 

\begin{equation}
    \dot{M} = 4 \pi r^2 \rho(r) v(r).
\end{equation}

\noindent where $v$ is the outflow speed. Following F18, we parameterize $v(r)$ as a beta-law,  

\begin{equation}
    v(r) = v_0 + (v_\infty - v_0) \displaystyle\left(1-\frac{R_\star}{r}\right)^\beta.
\end{equation}

\noindent where $\beta$ dictates the steepness of the acceleration, $v_\infty$ is the terminal wind velocity, and $v_0$ is the initial speed of the wind at the photosphere which must be non-zero to avoid the density tending to infinity at small $r$. Following F18, we adopt {\bf \mdot$=10^{-3}$\msunyr}, $\beta=3$, $v_\infty = 10$\kms, and a value of $v_0 = 0.0125$\kms\ is chosen to ensure that the density at the inner wind knits continuously with that at the outer photosphere at the location where $T(r) = T_{\rm eff}$. As with F18, the wind is propagated out to a distance $r_{\rm wind}$ = 20\rstar, though we compute a series of models at shorter $r_{\rm wind}$ to study the spectral evolution of the star throughout the launching of the wind. 

To compute the emerging flux through the wind, we must also specify a temperature profile $T(r)$. To compute $T(r)$, in this work we make the assumption that radiative equilibrium holds at all depths in the wind. To do this, we begin with the $T$ profile of the MARCS model, and at the point the wind meets the atmosphere we make the initial estimate of $T$ scaling as $r^{-0.5}$ from the inner to the outer extent of the wind. We then iterate on the $T$-structure to ensure that at every depth in the wind the condition, 

\begin{equation}
\int_{\lambda_1}^{\lambda_2} \kappa_\lambda (J_\lambda - B_\lambda) d\lambda = 0
\end{equation}

\noindent is satisfied. Here, $\kappa_\lambda$, $J_\lambda$ and $B_\lambda$ are the opacity, the intensity of the radiation field, and the Planck function respectively. To do this, we use {\sc Turbospectrum v19.1} to solve the radiative transfer equation between $\lambda_1 = 0.2$\um\ and $\lambda_2 = 12$\um, at wavelength sampling of $\lambda/\Delta\lambda = 500$, and determine $\kappa_\lambda$, $J_\lambda$ and $B_\lambda$ at all wavelengths and all depths. The wavelength limits $\lambda_1$ and $\lambda_2$ are chosen to encompass $>95$\% of the flux at all temperatures <6000K, which is the majority of the atmosphere and wind. In practice, iterating the $T$-structure results in very little shift from the initial estimated profile for those models where the superwind has yet to reach 2$R_\star$. For later models, a very small adjustment to the temperature profile was necessary, where $T(r/R_\star>2)$ had to be reduced by a few percent. After a small number of iterations the luminosity at every depth was consistent to within 5\%, and the luminosity at the outer and inner radii are consistent to within 0.05dex. Furthermore, very little change to the emergent spectrum was seen between the iterated and initial models.

\begin{figure}
    \centering
    \includegraphics[width=8.5cm]{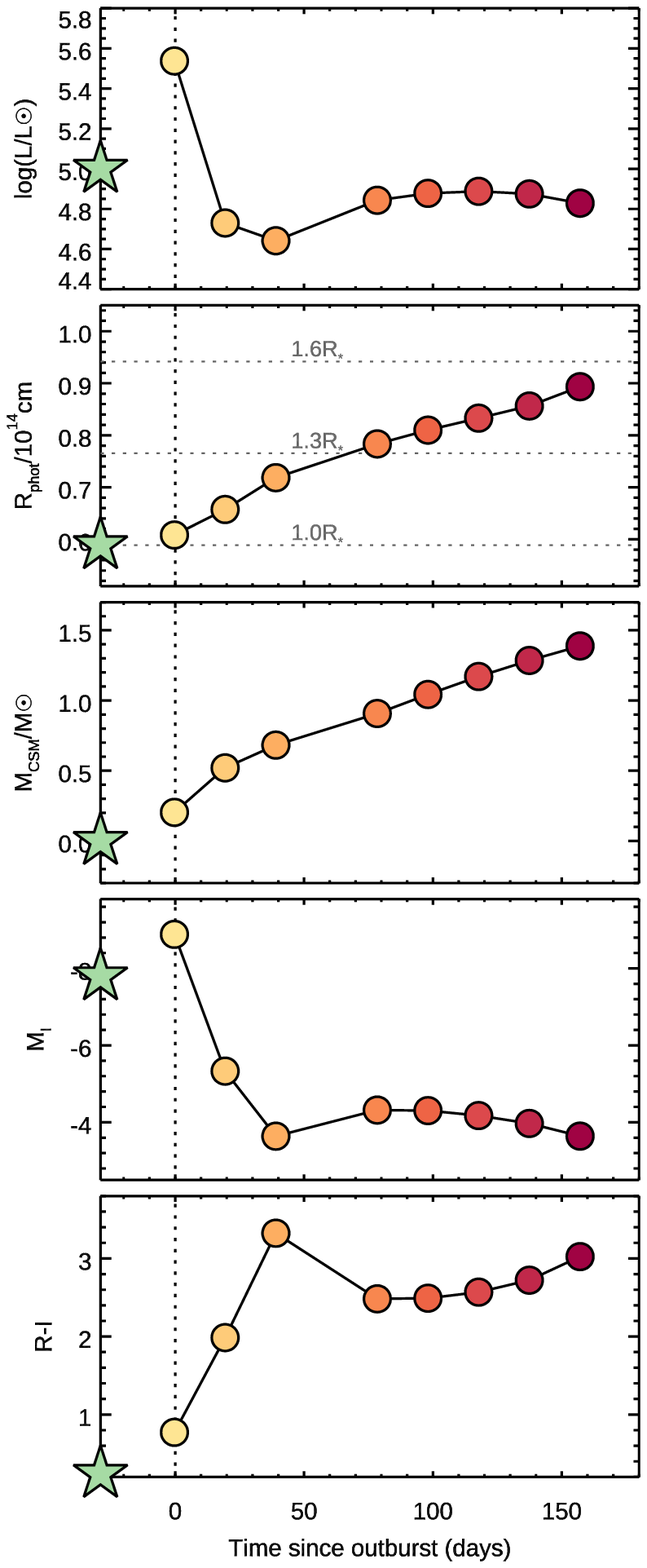}
    \caption{Evolution of the star's observables during the outburst. {\it Top}: bolometric luminosity; {\it second top}: the radius of the photosphere, defined as the flux-weighted average of the $R_\lambda (\tau=1)$ surface; {\it centre}: circumstellar mass, defined as the mass now above the original photosphere; {\it second-bottom}: absolute magnitude in the $I$-band; {\it bottom}: $R-I$ colour. The green star indicates the properties of the {\bf base} model. The vertical dashed line indicates the time at which the subsurface eruption breaks the surface of the star, defined in M20 as $t=140 $days, but defined here as $t=0$. }
    \label{fig:obs-mz}
\end{figure}

\begin{figure}
    \centering
    \includegraphics[width=8.5cm]{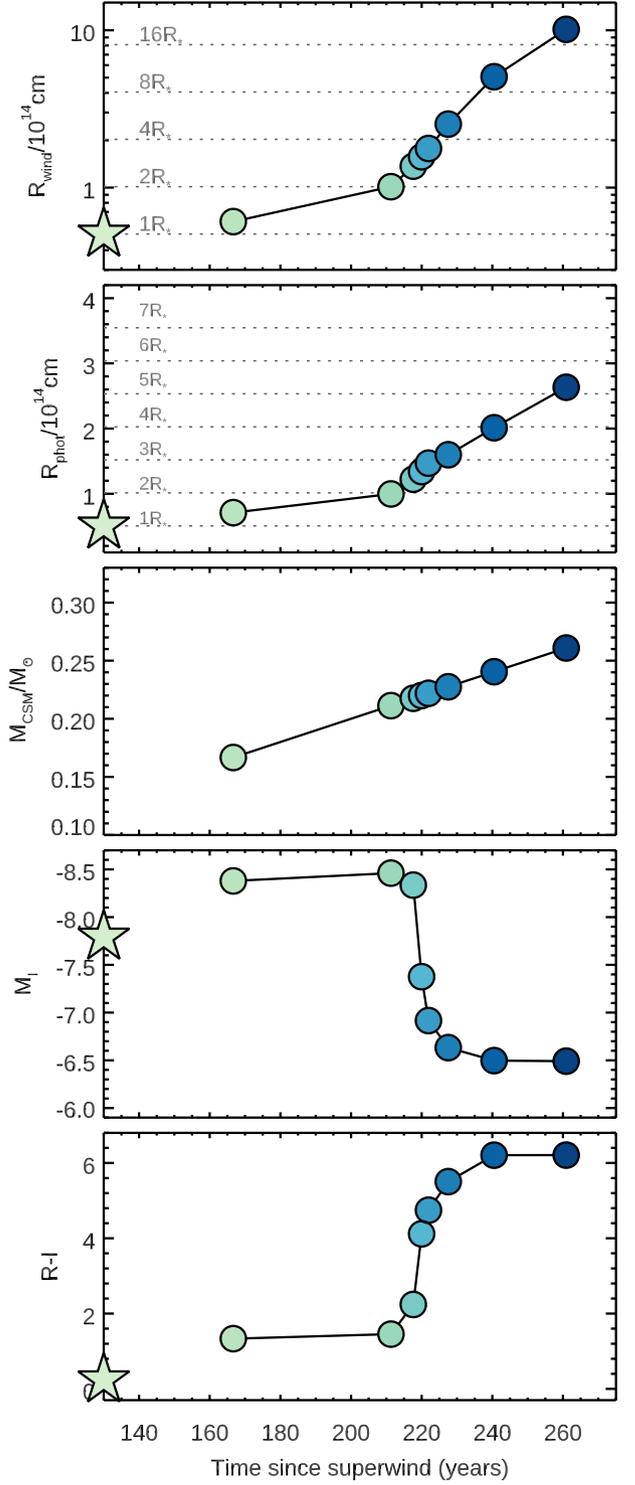}
    \caption{Evolution of observables during the superwind. The panels show the same as \fig{fig:obs-mz}, apart from the top panel which in this figure illustrates the radius of the outermost part of the wind. }
    \label{fig:obs-f18}
\end{figure}

\begin{figure*}
    \centering
    \includegraphics[width=\columnwidth]{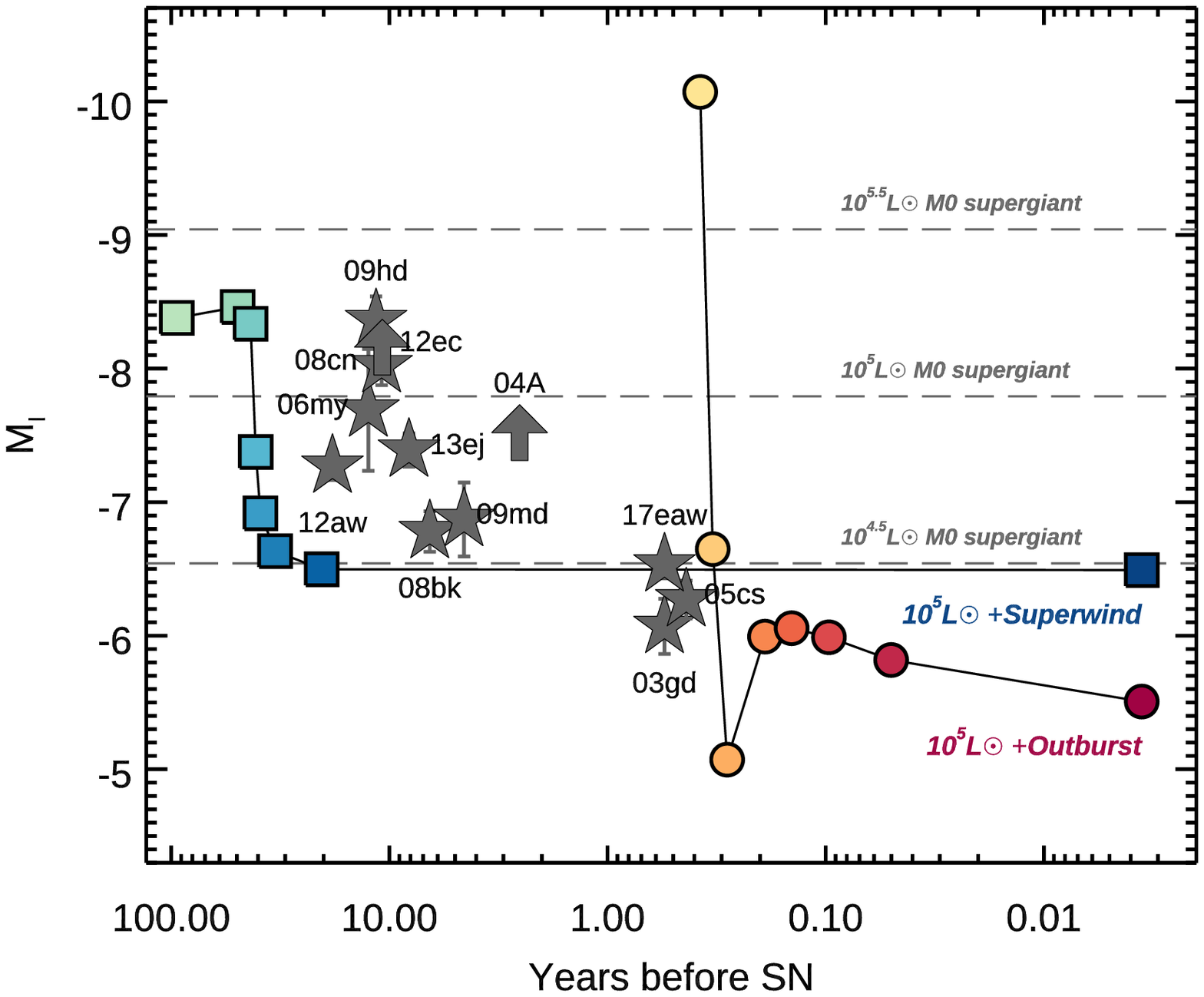}
    \includegraphics[width=\columnwidth]{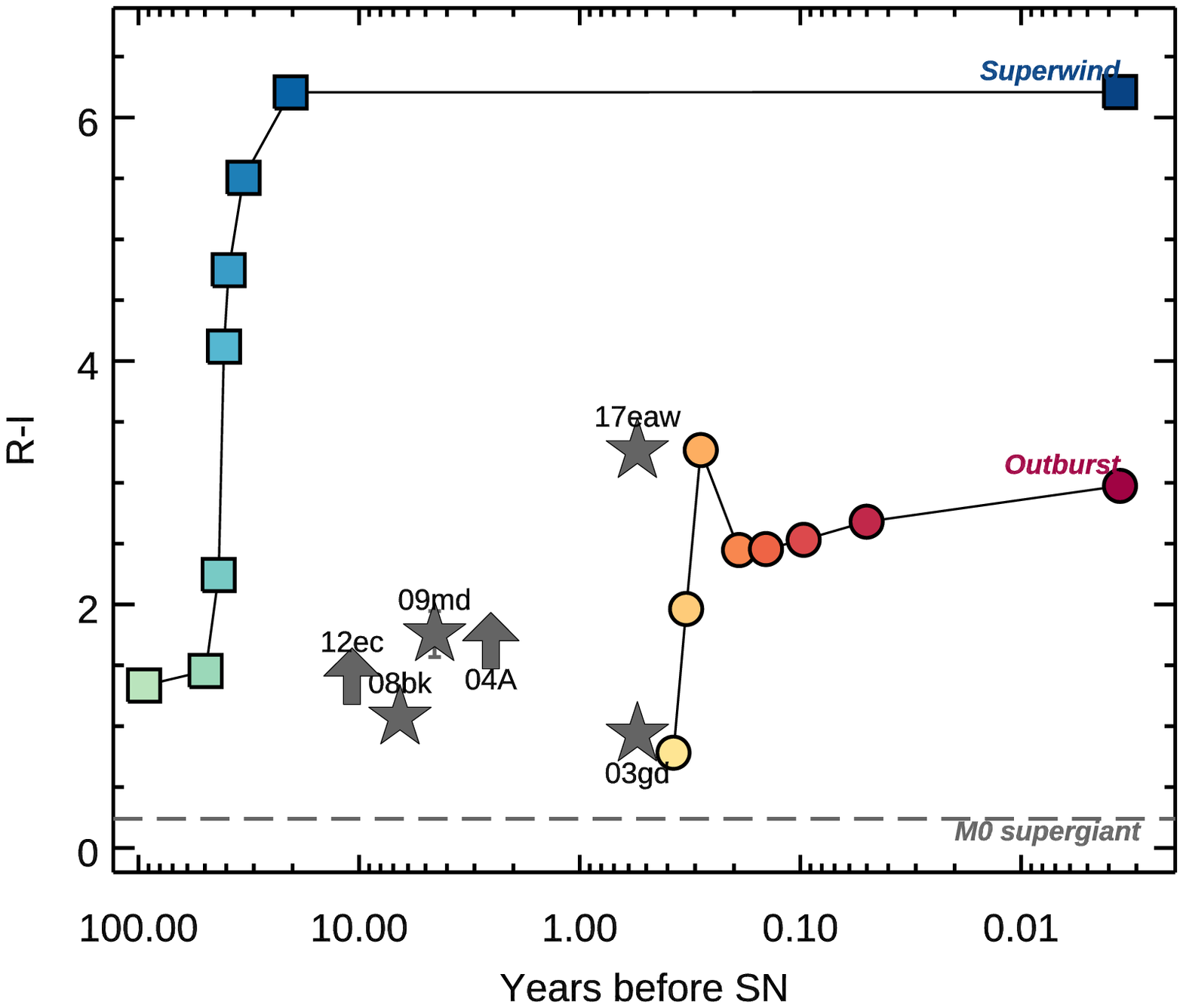}
    \caption{The $I$-band magnitude ({\it left}) and $R-I$ colour ({\it right}) as a function of time for each scenario tested. On each panel we overplot observations of SN progenitors, taking data from \citet{Davies-Beasor18,Davies-Beasor20} and references therein. We also overplot the location of an RSG of spectral type M0 for three different luminosities.}
    \label{fig:cmds}
\end{figure*}

\subsection{Computing the synthetic spectra}
To compute the emergent spectra of these models, we take the density and temperature profiles at each stage of the simulation for each of the two scenarios, and pass them through our modified version of {\sc Turbospectrum} \citep{Plez12,Davies-Plez21}. The code solves for the molecular equilibrium in the cooler layers, and computes the radiative transfer through the atmosphere under the assumption of local thermodynamic equilibrium. We assume Solar metallicity and Solar-scaled abundances in all spectra, noting that RSGs may display surface abundances consistent with mild CNO processing \citep{rsgcabund}. Though we do not know quantitatively what the impact of these approximations are, as we will see in the next Section the strengths of the molecular features are enormous, far beyond that seen in `normal' RSGs. Therefore, our conclusions are robust to these approximations provided they do not cause the TiO and ZrO band strengths to decrease by more than a factor of $\sim 5$, such is the effect on the progenitor colours (see Sect.\ \ref{sec:disc}).


\section{Results} \label{sec:spectra}
The synthesised spectra for the two scenarios are plotted in \fig{fig:spectra-mz} and \fig{fig:spectra-f18}, along with a 3800K {\sc marcs} model which was the progenitor. All spectra have been smoothed to a constant resolving power of $R \equiv \lambda/\Delta \lambda = 1000$ to make the spectral behaviour clearer around the molecular bands.

The time evolution of various physical properties and observables during the pre-SN events are illustrated in Figs.\ \ref{fig:obs-mz} and \ref{fig:obs-f18}. In each of these plots, the time-dependent photospheric radius $R_{\rm phot}$ (second-top panel in each figure) is defined as the flux-weighted average of the $\tau_\lambda = 1$ surface. The circumstellar mass $M_{\rm CSM}$ is defined as the mass above the {\it original} progenitor photosphere. 

Below, we discuss the results from each scenario in more detail.



\subsection{`Outburst' spectra}
The first model we compute is that shortly after the shockwave penetrates the surface layers, defined in M20 as $t=140$days, but defined here as $t=0$. The star expands slightly, and temperature at the surface increases by about 1200K, causing an increase in bolometric luminosity. The emergence of the shock leads to a `thickening' of the outer layers, effectively reducing the atmosphere scale height. These factors combined lead to a dramatic reduction in the molecular opacity, and an almost featureless spectrum aside from some weak atomic absorption that cannot be seen at the spectral resolution of Fig.\ \ref{fig:spectra-mz} {\bf ($R = 1000$)}. In terms of the star's behaviour in colour-magnitude space, the star briefly becomes bluer and brighter in the $I$-band. 

Only a few days after the shock breakout, the star surface cools to below the initial effective temperature, while the photosphere expands. Luminosity is not conserved during this time, and the star dims to become $\sim0.3$dex fainter than the initial model. This cooling of the photospheric layers, combined with the atmosphere becoming more extended, leads to an increase in molecular opacity. The TiO and ZrO bands in the optical, as well as the CN, CO and H$_2$O bands in the near-IR, all go into absorption. The cooler temperature, fainter luminosity and increased molecular opacity lead to the star becoming $\sim$4~mags fainter in the $I$-band, and $\sim 2$~mags redder in $R-I$, within a month of the breakout of the outburst. For the remaining $\sim$100 days of the simulation, the outer layers continue to expand, forcing the photosphere to larger radii. The photospheric temperature decreases very slightly during this time, causing the bolometric luminosity to remain constant to within $\sim$0.1dex, and the $R-$ and $I$-band fluxes to be stable to within a few tenths of a magnitude. According to M20, the model that best reproduces the early-time SN spectrum is the last of this series, meaning that the putative SN would occur $\sim$160 days after the emergence of the pre-SN outburst, which is the dark red spectrum in \fig{fig:spectra-mz} and the final datapoint in \fig{fig:obs-mz}.

Under this scenario, we can conclude that the signature of an impending SN would be a very brief brightening of a few days, during which the star became bluer, followed by a dramatic dimming and reddening over the next month, with the star remaining in a similar state for several more months until SN.  


\subsection{`Superwind' spectra}
During the first $\sim$160 years, the very shallow acceleration profile of the wind (set by the high value of $\beta$) leads to the CSM to accumulating close to the initial photosphere, causing the atmosphere to `thicken' as in the outburst scenario. As this happens, the continuum opacity increases, but the molecular line opacity stays roughly constant. The effect is to `fill-in' the absorption lines, leading to an SED which has a similar overall shape but has less absorption features. 

After $\sim$200 years the wind expands outwards beyond 2$R_\star$, and we begin to see the first molecular features as the CO bands go into emission, but in terms of its brightness at $R$ and $I$ the star has changed only very little. Just after 220 years, the outer `superwind' reaches 3$R_\star$, and here it approaches its maximum acceleration. It is at this point that we see the most dramatic change in the star's properties (see \fig{fig:obs-f18}), as the outer wind is now beginning to expand at a rate close to its terminal velocity. The very high mass-loss rate causes the effective flux-averaged radius to move outwards, reaching 5$R_\star$ after around 250 years. During this time, the optical absorption bands of TiO and ZrO become heavily saturated, the CO and CN bands in the $J$ and $H$ bands also go into absorption, while the CO and H$_2$O bands $>1.8$\um\ go into emission. This causes the star to become fainter in the $I$-band by $\sim$2\,mags, but the most dramatic effect is seen in the star's $R-I$ colour which becomes $\sim$ 6 magnitudes redder than the initial colour of the base model. 

At 20$R_\star$ (around 260 years after the onset of the `superwind') the local temperature drops below 1000K, at which point we would expect dust to form. Even the dustiest RSGs known, such as VY~CMa or WOH~G64 \citep[e.g.][]{Shenoy16,Goldman17} have mass-loss rates which are at least an order of magnitude below that of the superwind. Therefore, we would expect that once this star begins to form dust it will do so on a scale beyond seen in any field RSG. This would result in the star becoming even fainter and redder than our predictions presented here. 

The obvious aspect that differentiates this scenario from the `outburst' is the timescale involved. Whereas the time from outburst to SN is around 0.5 years, by contrast the `superwind' takes over 250 years to reach 20\rstar, which is the outer wind radius necessary to reproduce the observed features in the SN lightcurves \citep{Moriya17,Foerster18}. By the time of explosion, our results indicate that the star would've been exceptionally red for $>$40 years prior to core-collapse. The only way to make this CSM accumulate faster would be to increase \mdot\ and decrease $\beta$, at which point the `superwind' starts to become an analytic approximation of the `outburst' described in M20. 


\section{Comparison to observations of II-P progenitors} \label{sec:disc}
To compare our results to observations, we use the sample of II-P SNe from nearby galaxies presented in \citet{Davies-Beasor20}. This sample contains nearly all known II-P SNe with host galaxies within 30Mpc between 1999 and 2017, and so can be assumed to be volume limited. The only biases present in this sample are (a) any SNe that may have occurred in low surface brightness galaxies would likely be missing, owing to their host galaxy lacking pre-explosion imaging; and (b) a SN in a very crowded region --  namely SN~2016cok --- is missing from our sample, as the progenitor is impossible to identify without very late time observations with which to perform image subtraction \citep[see][]{Kochanek17}.

To compare these SNe to our results, we begin with the subset of objects which have pre-explosion detections or upper limits in the $F814W$ or $I$-band filters. For each of these objects, we determine the absolute $I$-band magnitudes from their pre-explosion photometry, their distance modulii, and extinction \citep[][ and refs therein]{Davies-Beasor18,Davies-Beasor20}. We also define the time before SN that each progenitor was observed as being the difference between the observation date and the SN discovery date. From this subset of progenitors, we next identify those objects which have colour information in the form of a related detection in the $R$ (or $F606W$) band. The magnitudes and colours of the II-P progenitors as a function of the number of years before SN that they were observed are plotted in \fig{fig:cmds}. In total, we have 12 events with pre-explosion $I$-band imaging, and 6 events with a constraint on $R-I$ \footnote{We chose the $R-I$ colour for this study as it leads to the largest homogeneous sample of progenitors. There are a small number of progenitors with additional pre-explosion detections or limits in the $V$-band, but these data do not add anything conclusive to the discussion here.}. 

On each panel of \fig{fig:cmds}, we overplot the predicted time evolution of the two scenarios studied in the previous section. For the `outburst' the SN time is assumed to be 160 days after the outburst; and in the `superwind' model it is taken to be {\rm 260 years} -- the time at which the wind reaches 20\rstar. For each scenario, the base input model was a $10^5$\lsun\ star with spectral type M0. The locations of this input star on these diagrams is marked in grey, as well as the locations of similar stars with different luminosity (in the left-hand panel only).  

The first thing we note here is that no RSG SN progenitor has yet been observed {\it after} we would expect the outburst to have happened in the `outburst' scenario. Though three SNe (03gd, 05cs and 17eaw) were observed $\sim$200 days before explosion, which is roughly when the outburst would be expected, we are currently unable to rule out the `outburst' scenario from these data. 

However, we are able to make conclusive arguments regarding the `superwind' scenario. Recall that in this scenario, the pre-explosion CSM accumulates slowly over several decades, causing the star to become a factor of 3 fainter in $I$ and over 5 mags redder in $R-I$ $> 40$years before explosion. This is strongly contradicted by observations of nearby SNe.

In \fig{fig:cmds}, we first see that none of the progenitors were as faint as we would expect a $10^5$\lsun\ progenitor with a superwind to be. Furthermore, of the four objects for which we have pre-explosion colours (right-panel of \fig{fig:cmds}), three had colours of a normal M0 supergiant. The one outlier is 17eaw, which is still several magnitudes bluer than we would expect if it were experiencing a superwind; we discuss this object in more detail below. Zero out of four objects is obviously inconsistent with the conclusions of F18 that 92\% of II-P SNe progenitors experienced superwinds before they exploded. Formally, the probability that we would randomly find zero objects with superwinds in an unbiased sample of four progenitors, if the intrinsic fraction was 92\% (as reported in F18), is $p \simeq 5\times10^{-5}$. 

\subsection{The case of SN 2017eaw}
This SN is the one object in our sample that could vaguely be described as displaying a red colour of order our model predictions. As such, we afford this object some further discussion. 

The progenitor underwent a modest increase in brightness of 20\% at 4.5\um, around 3 years before SN. Meanwhile, it remained stable at 3.6\um\ \citep{Kilpatrick-Foley18,vanDyk19,Rui19}, and displayed no detectable variability in these bands in the 50-500 days before explosion. It showed little evidence of variability at $I$ for the 12 years preceding explosion. Finally, \citet{Johnson18} claimed that the progenitor's flux was stable to within $\pm$0.4\% in the $R$-band across multiple observations between 0.5-9 years. However, these authors also point out that they do not detect the progenitor in any individual observation, therefore we treat this evidence with caution. 

Each of \citet{Kilpatrick-Foley18}, \citet{vanDyk19} and \citet{Rui19}  argue for a very cool effective temperature {\it and} substantial circumstellar dust, based on the progenitor's spectral energy distribution. Specifically, the progenitor was redder at $R-I$ and brighter at 4.5\um\ than is expected from standard {\sc phoenix} or {\sc marcs} models; a phenomenon that both works attributed to dust. However, it was shown in \citet{Davies-Plez21} that attaching a modest wind to a 3800K (=M0) model atmosphere can adequately explain this behaviour. At mass-loss rates of $0.3-1.0\times 10^{-5}$\msunyr, the wind increases the TiO absorption mimicking a much later spectral type and a redder $R-I$, whilst pushing the CO bands at $\ga$4\um\ into emission to reproduce the observed increase in flux at 4.5\um. 

Taking the above into account, we conclude that SN~2017eaw did {\it not} undergo an outburst or a superwind-like event prior to exploding. Rather, the red pre-explosion colour and mild brightening at 4.5\um\ are instead better explained by a modest increase in mass-loss rate to around $\sim 10^{-5}$\msunyr; several orders of magnitude shy of being enough to generate $\sim$0.1\msun\ of CSM by the time of SN. Of course, we cannot exclude that the progenitor underwent some further, more substantial variability in the $\sim$6 months between the final progenitor observations and the SN.

\subsection{Possibility of sample biases}
Finally, we discuss the possibility that the local sample of SNe used here to critique the predictions of each scenario is biased with respect to those works which claim near-ubiquity for {\rm greatly elevated CSM densities} at the point of explosion; specifically those of M18 and F18. 

The objects in F18 are from a magnitude limited survey, in that the sample contains all II-P SNe that were detected during an observing campaign. Though systematic, the F18 sample is likely biased toward brighter SNe. The sample of SNe in M18 on the other hand is more heterogeneous, in that the data for these objects were taken from the literature, and chosen for their well-sampled lightcurves in terms of both time and multi-band coverage. Finally, the data we present to compare to the synthetic spectra of the two scenarios is nominally volume limited, though there is potentially a bias towards SNe in bright (and therefore metal rich) galaxies which are more likely to have archival HST observations. 

Despite their contrasting approaches to drawing their samples, each of the M18 and F18 studies concluded the same: that the majority of II-P lightcurves are consistent with $\ga$0.1\msun\ of CSM at the point of explosion (13/20 and 35/37 objects for M18 and F18 respectively). Therefore, no matter how the sample of SNe is compiled, the results are qualitatively the same: most II-P SNe display evidence of atypically large amounts of CSM at explosion, irrespective of how the sample is compiled.

Furthermore, we can check that our SN progenitor sample is not biased with respect to the M18 sample since there are several objects in common. Specifically, four SNe from M18 (05cs, 12aw, 12ec, 13ej) have progenitor detections in the $I$-band. Three out of these four were determined to have $\ga$0.2\msun\ of CSM at explosion (the exception being 05cs, which according to M18 had $<$0.03\msun). Despite this, the pre-explosion brightnesses of these three SN were all consistent with an unobscured progenitor with $L \sim 10^5$\lsun\ \footnote{Unfortunately, only one of these progenitors (12ec) has pre-explosion imaging in the $R$-band, in which the object is not detected, with an upper limit that has little diagnostic power.}. Though the number of overlapping objects is small, we conclude that the apparently `normal' brightness and colours of the observed progenitors with respect to the predictions of the `superwind' scenario is unlikely to be an artefact of sample biases. 



\section{Conclusions} \label{sec:conc}
Several recent studies in the literature have claimed that abnormally dense circumstellar material (CSM) is almost always present around red supergiants (RSGs) at the point of collapse. In this paper we have tested two scenarios under which such large amounts of CSM may be generated: a short `outburst' lasting less than a year, with a mass-loss rate of \mdot$\ga0.1$\msunyr{}; and an extended `superwind' phase with \mdot$\sim 10^{-3}$\msunyr\ lasting $\sim$100 years. Our conclusions may be summarised as follows:

\begin{itemize}
    \item Each scenario creates large amounts of optically thick circumstellar gas, which attenuates the flux in the optical by several orders of magnitude. This means that a RSG on the verge of core-collapse would be very red and very faint in the optical, regardless of how or on what timescale the material was produced. 
    \item The much slower build-up of CSM within the `superwind' scenario causes the progenitor to be conspicuously red and faint for many decades preceding supernova. This is strongly inconsistent with observations of nearby SNe, which have always appeared to have colours and magnitudes normal for mid- to late- M supergiants. 
    \item Therefore, whatever the mechanism for generating this CSM, it must do it on a very rapid timescale. Specifically, the build-up of the CSM must happen within a year of core-collapse.
\end{itemize}

The final overarching conclusion we can make from this work is that, shortly before core-collapse, RSGs must undergo some prodigious mass-losing event which radically alters the appearance of the star. Therefore, the signature of an imminent explosion should be a dramatic change in the progenitor stars' optical -- near-IR photometry on timescales of less than a month. Such a signature should be detectable in the coming era of wide-field short cadence photometry.

\section*{Acknowledgements}
We thank Viktoria Morozova for providing the models for the `outburst' scenario, and for discussions during the course of this work. We also thank Emma Beasor and Nathan Smith for useful discussions, and the anonymous referee who's careful review helped us improve the paper.  

\section*{Data Availability}
The synthetic spectra generated as part of this study are available on request to the authors.



\bibliographystyle{mnras}
\bibliography{biblio} 



\bsp	
\label{lastpage}
\end{document}